\documentclass[aps,prl,reprint,10pt,superscriptaddress,amssymb,amsfonts,footinbib]{revtex4-1}

\usepackage[caption=false]{subfig}
\usepackage{amsmath}
\usepackage{amssymb}
\usepackage{graphicx,bm}
\usepackage{verbatim}
\usepackage{mathrsfs}
\usepackage{color}
\usepackage{floatrow}
\usepackage{dcolumn}
\usepackage{amssymb}
\usepackage[normalem]{ulem}
\usepackage{wasysym}
\usepackage{hyperref}
\usepackage{soul,xcolor}
\usepackage{lineno}
\begin{document}

\title{
Bipolaron liquids at strong Peierls electron-phonon couplings
}

\author{Alberto Nocera}\thanks{These two authors contributed equally}\email{alberto.nocera@ubc.ca}
\affiliation{Department of Physics and Astronomy, University of British Columbia, Vancouver, British Columbia, V6T 1Z1 Canada}
\affiliation{Stewart Blusson Quantum Matter Institute, University of British Columbia, Vancouver, British Columbia, V6T 1Z4 Canada}

\author{John Sous}\thanks{These two authors contributed equally} \email{js5530@columbia.edu}
\affiliation{Department of Physics, Columbia University, New York, New York 10027, USA}

\author{Adrian E. Feiguin}
\affiliation{Department of Physics, Northeastern University, Boston, Massachusetts 02115, USA}

\author{Mona Berciu}
\affiliation{Department of Physics and Astronomy, University of British Columbia, Vancouver, British Columbia, V6T 1Z1 Canada}
\affiliation{Stewart Blusson Quantum Matter Institute, University of British Columbia, Vancouver, British Columbia, V6T 1Z4 Canada}

\date{\today}

\begin{abstract}
  We use the Density Matrix Renormalization Group method to study a one-dimensional chain with Peierls electron-phonon coupling, which describes  the modulation of the electron hopping by lattice distortions.  We demonstrate that this system is stable against phase separation in the dilute density limit. We only find phase separation numerically for  large couplings for which  the linear approximation for the electron-phonon coupling becomes invalid; this behavior can be stabilized in a narrow sliver of the physical parameter space if the dispersion of the phonons is carefully tuned.
These results indicate  that in the dilute electron density limit, Peierls bipolaron liquids  are generically stable, unlike in other models of electron-phonon coupling.  We show that this behavior extends to finite carrier concentrations of up to quarter filling. This stability of low-density, light-mass bipolaron liquids in the Peierls model opens a path to high-$T_c$ superconductivity 
  based on a bipolaronic mechanism, in higher dimensions.
\end{abstract}

\maketitle

\normalsize

{\em Introduction.---} A primary goal in condensed matter physics
targets the discovery and understanding of unusual  phases of matter
that arise from  strong correlations. Correlated
electron-lattice systems manifest in fascinating ways in many experimentally relevant situations, such as
polaronic phenomena in the dilute electron density limit \cite{BipolaronsReview,MillisPolaronFL,Dominic,LightBP,SousBondP}, and charge order and superconductivity at finite electron concentrations \cite{BCS2, ScaletterScalapino, KivelsonHolstein2}.
In particular, the search for alternative 
microscopic mechanisms for high-$T_c$ superconductivity has stimulated a renewed interest 
in the study of bipolarons -- bound states of two polarons, where a polaron is an electron dressed by a cloud of phonons -- in the presence of different forms of electron-lattice coupling.

Recent work~\cite{LightBP} has in fact demonstrated that the Peierls electron-phonon coupling~\cite{SSH1a,SSH1b,SSH1c}, which describes the modulation of the electron hopping due to lattice
distortions, gives rise to strongly bound but light-mass bipolarons. As a result, these
should remain phase coherent up to high temperatures, opening a
possible route to phonon-mediated high-$T_c$ superconductivity whose 
fingerprints may have already been observed experimentally in the material Ba$_{1-x}$K$_x$BiO$_3$~\cite{Plumb, Johnston}. A possible hindrance to this scenario are competing
instabilities that could favor a different order. For example, for the
more studied generalized Holstein~\cite{Holstein1, Holstein2} 
and Fr\"ohlich~\cite{Froh1,Froh2} models
in which the lattice distortion modulates the electron's on-site energy, phase separation sets in at fairly weak electron-phonon couplings for low carrier concentrations \cite{Boncaextendedbipolaron,LiebFroh,MonodeepPS}. This is because bipolarons in these models become increasingly heavy
at stronger couplings \cite{ProkofievFrohPolaron1,Kornilovitchpolarongeneral1,Boncabipolaron} and experience a phonon-mediated  long-range electron-electron
attraction. The latter favors clustering of bipolarons  in order to minimize the total energy, thus  inducing phase separation.  This phenomenon presents a major obstacle to bipolaronic superconductivity (at any temperature) in these models.  In contrast, as stated earlier, the Peierls coupling favors light
bipolarons even at very large couplings because it mediates effective
{\em pair-hopping}  interactions that enhance the kinetic energy of
bipolarons \cite{LightBP}. As such, it is unclear whether phase separation occurs in the Peierls model at finite electronic densities. In this regime, the lack of controllable analytical approaches calls for the use of unbiased numerical techniques.

\begin{figure}[h!]
\centering
\hspace{0.0cm}\includegraphics[width=0.95\columnwidth]{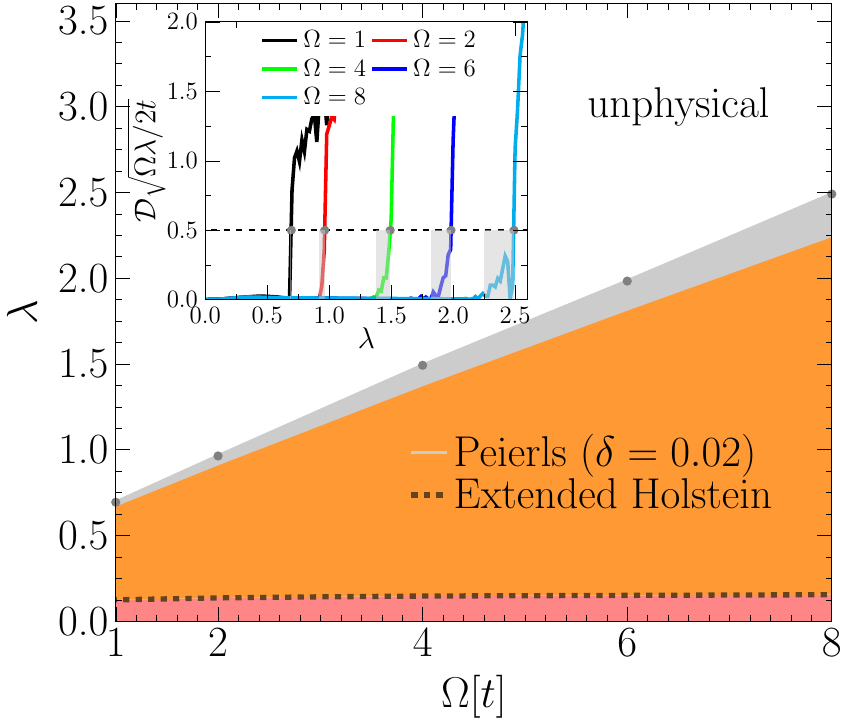}
	  \caption{(color online) Main panel: Phase diagram of the 1D Peierls model
	  contrasted against that of  the extended Holstein (EH) model in the dilute electron density limit. 
	  The latter represents a lattice model that mimics the physics of the Fr\"ohlich coupling, 
	  see Ref. \cite{Boncaextendedbipolaron}, for which $\lambda =  g^2/(2\Omega t)$.
	  The phase boundary defined by $\lambda_c(\Omega)$ separates a 
      stable liquid of bipolarons from phase separation. For the Peierls model, the bipolaron liquid is
      stable for $\lambda < \lambda_c$ (orange region), becomes unstable to phase separation for 
      $\lambda_c <\lambda < \lambda_{s}$ (grey region), 
      and the latter quickly gives way to an unphysical regime for $\lambda > \lambda_{s}$ 
      (white region), see text for details.
      For the EH model, the dashed line labels critical values $\lambda_c(\Omega)$ separating a
      stable liquid of bipolarons for $\lambda < \lambda_c$ (red-orange region) from the phase separation for
	 $\lambda > \lambda_c$ (all regions above).
	 In clear contrast to the EH model, in the Peierls model
  the bipolaron liquid is stable up to much larger
  $\lambda$, with $\lambda_c\rightarrow \infty$ as $\Omega\rightarrow\infty$. 
  These DMRG results are for a system with $L=32$ sites, $N=6$
  electrons and using a finite dispersion parameter $\delta=0.02$.
	  Inset: Staggered displacement amplitude ${\cal D}$, defined in Eq.~(\ref{eq:StagDisp}),
	  as a function of $\lambda$ for different values of  $\Omega$ in the Peierls model. 
	  The horizontal dashed line at 1/2 defines the limit beyond which phonon displacements become unphysical.}
	  \label{fig:PhaseDiagram}
 \end{figure}

\begin{figure}[t]
  \centering
\hspace{-0.3cm} \vspace{-0.7cm}
\includegraphics[width=1.07\columnwidth]{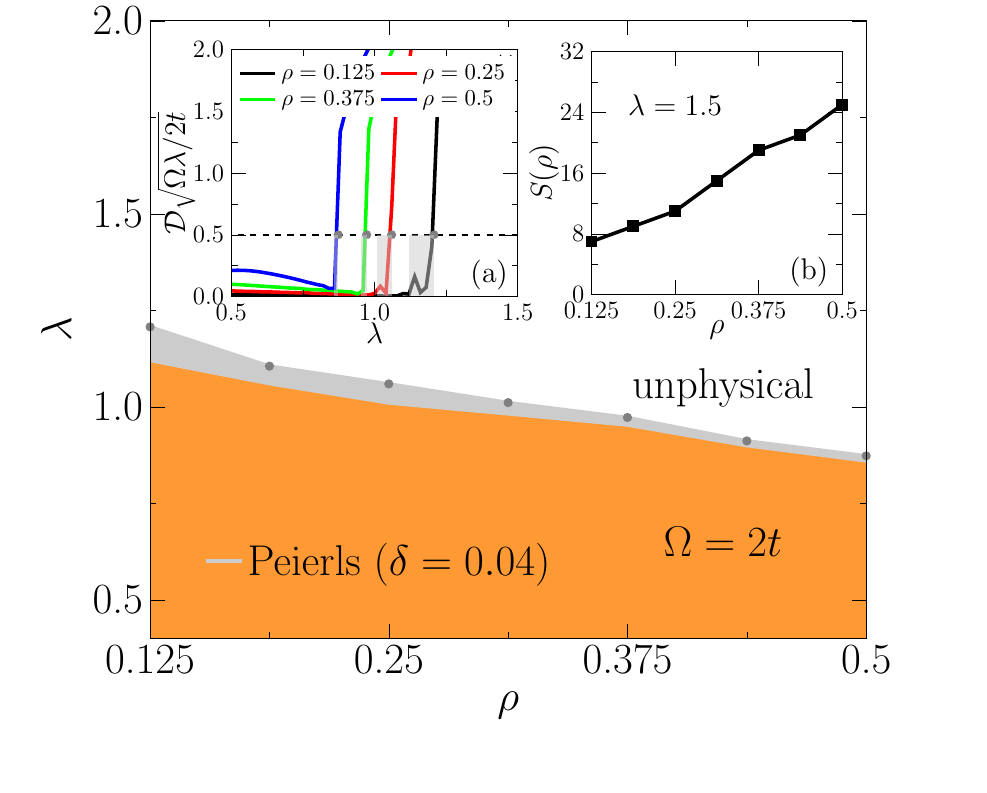}
	  \caption{(color online) 
	  Main panel: Phase diagram of the 1D Peierls model as a function of  the carrier filling $\rho=N/L$ 
	  ($\rho = 0.5$ corresponds to quarter filling) and $\lambda$ from DMRG for a system with $L=32$ sites and $\Omega=2t$.
	  Inset (a): Staggered displacement amplitude (Eq.~\eqref{eq:StagDisp}) as a function of $\lambda$ for various $\rho$. The horizontal dashed line at $1/2$ sets the  limit of physicality for the phonon displacements.
	  Inset (b): Size of the dimerized region in the phase separated phase as a function of $\rho$.
	  }
 \label{fig:PhaseDiagram2}
 \end{figure}

In this Letter we use the Density Matrix Renormalization Group (DMRG) method \cite{WhiteDMRG} to show 
that in one dimension (1D), a
dilute liquid of bipolarons is stable against phase separation up to strong Peierls  electron-phonon couplings. Phase separation occurs only when lattice distortions are so large such that the linear approximation breaks down and the model becomes unphysical. Refinements  such as the inclusion of a finite phonon dispersion may render phase separation possible for sufficiently large couplings, but also pushes the critical coupling to larger values. We illustrate this main result in Fig.~\ref{fig:PhaseDiagram} where we contrast the stability of a liquid of Peierls bipolarons with the behavior of a generalized extended Holstein model in which phase separation sets in at very low couplings. We also confirm that this phenomenology extends to finite electron concentrations up to quarter filling (see Fig.~\ref{fig:PhaseDiagram2}).

Instability of polarons or bipolarons to phase separation operates more effectively in lower dimensions.  Therefore, given our results in 1D, we expect the stability of a liquid of Peierls bipolarons to extend to higher dimensions.  This, coupled with the arguments of Ref.~\cite{LightBP} showing that the Peierls model realizes light-mass bipolarons, further supports the possibility of high-$T_c$
superconductivity based on a bipolaronic mechanism in physically relevant dimensions~\cite{zhang2021peierls,Lim2021}.

{\em Model.---} We consider the one-dimensional Peierls model of electron-phonon coupling~\cite{SSH1a,SSH1b,SSH1c,SSH2a,SSH2b} 
$${\mathcal H} = {\mathcal H}_{\rm e} + {\mathcal{H}}_{\rm ph} + V_{\rm
  e-ph}.$$ 
Here $${\mathcal H}_{\rm e} = - t \sum_{i,\sigma} c_{i,\sigma}^\dagger c_{i+1,\sigma} +
{\rm H.c.}$$ describes nearest-neighbor hopping of electrons of
spin  $\sigma \in \{\uparrow, \downarrow\}$ in a single electronic band with creation operator $c_{i,\sigma}^\dagger$ at site $i\in\{1,..,L\}$, and number operator $\hat{n}_i = \sum_\sigma \hat{n}_{i,\sigma}$. The Peierls electron-phonon coupling describes the modulation of the hopping integral due to lattice distortions in the linear approximation:
\begin{align*}
V_{\rm e-ph} &= g \sum_{i,\sigma} ( c_{i,\sigma}^\dagger c_{i+1,\sigma} + {\rm H.c.} ) (b_i^\dagger +
b_i - b_{i+1}^\dagger -b_{i+1}). \nonumber
\end{align*}
Phonons belong to an optical Einstein mode of frequency $\Omega$ (we set $\hbar =
1)$ with $$\mathcal{H}_{\rm ph} = \Omega \sum_{i} b_i^\dagger b_i + \delta\frac{\Omega}{4}\sum_{i}(b_i^\dagger+b_i-b_{i+1}^\dagger-b_{i+1})^2.$$ Here, a phonon is  described by a boson creation
operator $b_i^\dagger$ at site $i\in \{1,..,L\}$.  The $\delta$ term gives a dispersion to the optical phonons (we set the lattice spacing $a=1$ and oscillator mass $M=1$): $\Omega_q=\Omega\sqrt{1+4\delta\sin^2(q/2)}$ shows hardening at the zone boundary $q=\pm \pi$ if $\delta\ne0$.  We characterize the strength of Peierls electron-phonon coupling via the dimensionless coupling $\lambda = {2g^2}/{(\Omega t)}$.

{\em Methods.---} We study $N\le 16$ electrons in the zero magnetization sector $S^z_{\rm Tot} = 0$ of the 
Peierls model on chains of  lengths  $L\le 48$. We use DMRG to compute the ground state, utilizing up to 
$n_{{\rm ph}_{\rm max}} + 1 = 20$ phonon states to represent the local phonon Hilbert space.
In the literature, DMRG has been often used to study one dimensional correlated Hamiltonians with Holstein electron-phonon coupling, seldomly with Peierls electron-phonon coupling~\cite{DMRGE1,DMRGE2,DMRGE3,DMRGE4,DMRGE5,DMRGE6,DMRGE7,DMRGE8,DMRGE9,DMRGE10,DMRGE11,DMRGE12,DMRGE13,DMRGE14,DMRGJS}.
Our numerical results were converged  with respect to the bond dimension $m$.  A maximum $m=600$ provides 
convergence with a truncation error smaller than $5\times10^{-7}$ for open boundary conditions 
(OBC) and $5\times10^{-6}$ for periodic boundary conditions (PBC), 
see the Supplemental Materials for more information~\cite{SupplementaryMaterials}. 

Note that unlike in Quantum Monte Carlo approaches, where an explicit cutoff on the amplitude of the lattice distortions is 
introduced to avoid unphysical changes in the sign of the hopping term~\cite{FradkinHirsch,AssaadHohen} or a restricted interval of interaction strengths is explored, in DMRG simulations one may use the $\delta$ phonon term as a Lagrange multiplier that energetically penalizes  unphysical changes in the sign of the hopping, {\it i.e.}, as a physical constraint on the length of the bonds.

  \begin{figure}[t]
  \centering
\includegraphics[width=0.95\columnwidth]{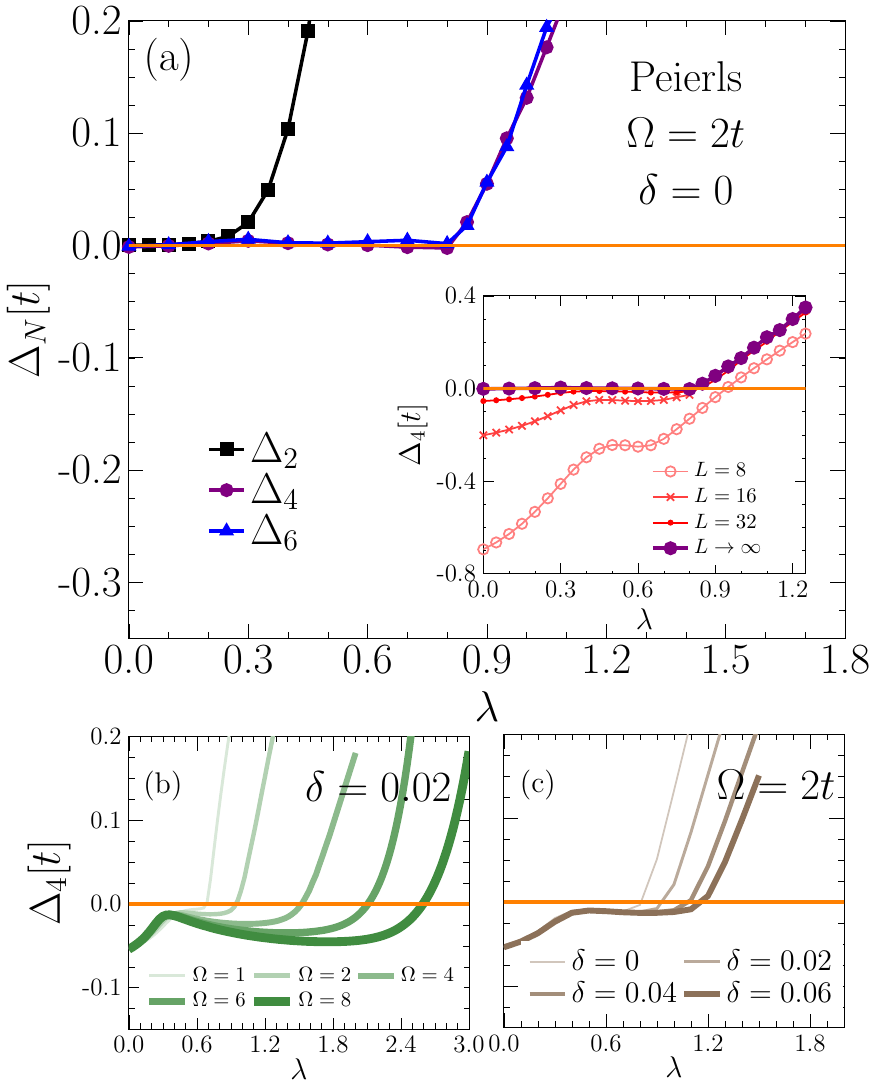}
	  \caption{(color online) (a) Main panel: Energies $\Delta_2=2E_1-E_2, \Delta_4=2E_2-E_4, \Delta_6=E_2+E_4-E_6$ against $\lambda$ for $\Omega=2t$, where $E_N$ is the GS energy for $N$ electrons on a chain of infinite length. For $\lambda > \lambda_c \approx 0.8$, $\Delta_4>0$ and $\Delta_6>0$ signal that the $N$ carriers bind, indicating phase separation (see text for more details). Inset: Scaling of $\Delta_4$ with system size $L$ proves that for $\lambda < \lambda_c\approx 0.8$, $\Delta_4\rightarrow 0$ as $L\rightarrow \infty$. The scaling was performed using data up to $L=64$ (not shown) and assuming $\Delta_4(L)=a_0+\frac{a_2}{L^2}+\frac{a_4}{L^4}$ for $\lambda < \lambda_c\approx 0.8$ and  $\Delta_4(L)=a_0+\frac{a_2}{L}$ for $\lambda > \lambda_c\approx 0.8$. } (b) $\Delta_4$ as a function of $\lambda$ for various values of the phonon frequency $\Omega$, at $\delta=0.02$. (c) $\Delta_4$ as a function of $\lambda$ for various values of  $\delta$ of the phonon dispersion, at $\Omega=2t$.
 \label{fig:BoundStates}
 \end{figure}

{\em Dilute electron density limit.---}  
Previous work showed that Peierls bipolarons are  stable against dissociation into single 
polarons for all $\lambda$, unless an extremely large Hubbard repulsion is present (a possibility ignored 
in this work) \cite{LightBP}. We first analyze the stability of a dilute liquid of these Peierls bipolarons.  
We use DMRG to find the ground-state (GS) energies $E_N$ for $N=1,2,4,6$ electrons on a chain with $L=32$ sites shown in ~Fig.~\ref{fig:BoundStates}.
(We have performed finite-size scaling of the results confirming that $L=32$ data is representative of the 
infinite chain limit.) We define $\Delta_2=2E_1-E_2,\Delta_4=2E_2-E_4, \Delta_6=E_2+E_4-E_6$ and study their 
dependence on $\lambda$ in Fig.~\ref{fig:BoundStates}(a). In the thermodynamic limit, all 
$\Delta_N\ge 0 $: $\Delta_N=0$ indicates that smaller complexes, each composed of fewer than $N$ particles, are 
energetically favorable ({
\it e.g.}, $\Delta_4=0$ means that the $N=4$ GS consists of two bipolarons), while 
$\Delta_N >0$ implies that bound state of the $N$ carriers is more stable. Figure \ref{fig:BoundStates}(a) shows that $\Delta_2>0$ for 
all $\lambda$, confirming that the $N=2$ GS always corresponds to a bipolaron, 
in agreement with \cite{LightBP}.
Both $\Delta_4$ and $\Delta_6$ only become positive above roughly 
the same $\lambda_c\gtrsim 0.8$, showing the tendency of all carriers present in the system to coalesce 
if $\lambda \ge \lambda_c$. (We note that negative $\Delta_2,\Delta_4$ values for $\lambda < \lambda_c$ 
are due to finite-size effects, see the inset of Fig.~\ref{fig:BoundStates}(a).  Additional analysis 
based on Maxwell construction is presented in the Supplemental Materials~\cite{SupplementaryMaterials}). Note that below $\lambda_c$ we find $E_{2N}=N E_2$, and $E_2$ agrees with the single bipolaron energy of Ref. ~[\onlinecite{LightBP}]. This confirms that this dilute bipolaron liquid is formed of essentially isolated bipolarons whose properties follow from Ref. [\onlinecite{LightBP}]; in other words they continue to be of light-mass and strongly bound in the dilute liquid phase.

Figure~\ref{fig:BoundStates}(b) shows the evolution of $\lambda_c$ with $\Omega$, which we use to identify the location of the phase boundary between the bipolaron liquid and  phase separation in the  phase diagram of Fig.~\ref{fig:PhaseDiagram}. 
Figure~\ref{fig:BoundStates}(c) shows that  $\lambda_c$ drifts to larger values with increasing $\delta$, indicating that the instability of bipolarons to clumping (phase separation) is unlikely in realistic systems (see Supplemental Materials for an extended discussion).  This drift occurs due to a competition before the onset of phase separation between the $\delta$ term, disfavoring dimerization, and the electron-phonon interaction, which drives dimerization at large couplings.   Special combinations  of $\delta$ and $\lambda$ render phase separation possible within the domain of physicality of the model, as we discuss next.

  \begin{figure}[t]
  \centering
\hspace{-0.35cm}\includegraphics[width=1.025\columnwidth]{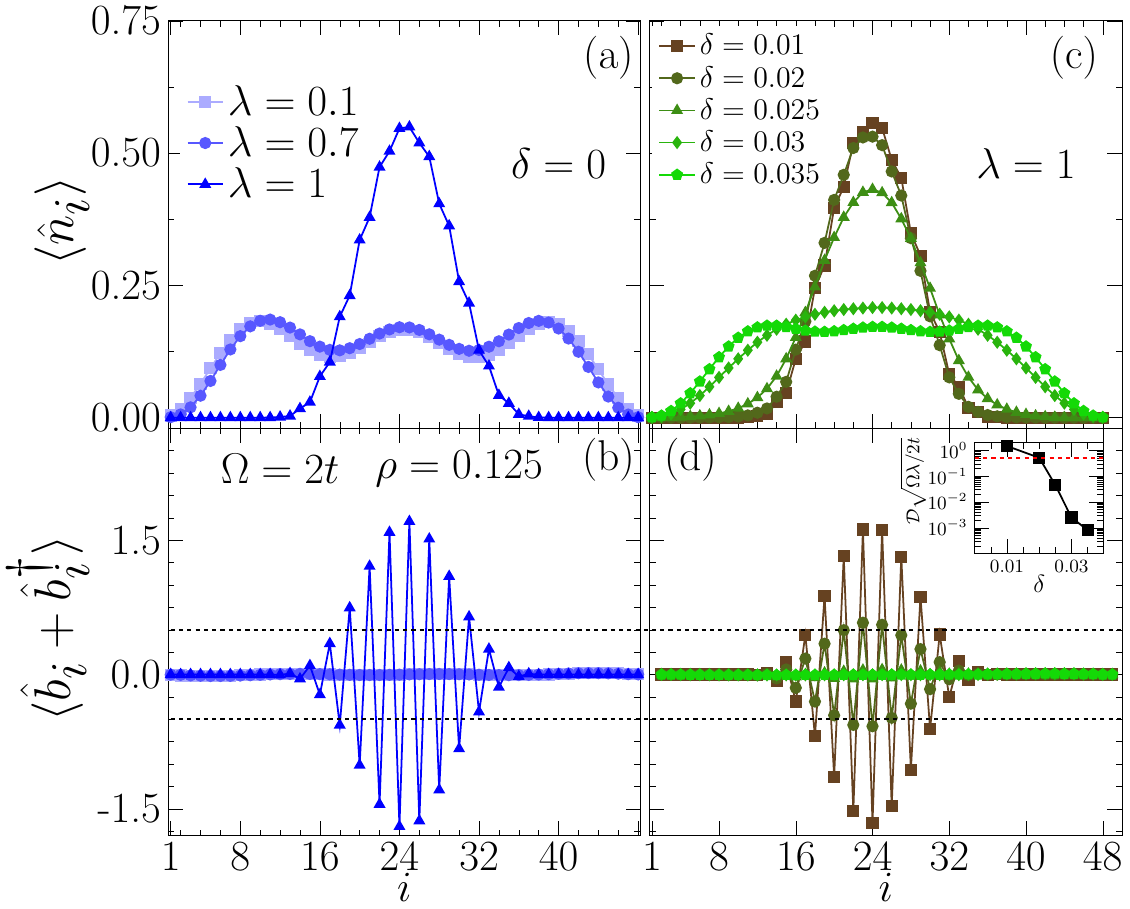}
	  \caption{(color online) GS expectation values of the  average occupation number 
	  $\langle \hat{n}_i\rangle $ ((a) and (c)),  lattice distortion $\langle b_i+b_i^\dagger\rangle$ ((b) and (d)). 
	  The results were obtained for a chain with $L=48$ and OBC and with $N=6$ electrons ($\rho=0.125$) and 
	  at phonon frequency $\Omega=2t$. Panels (a) and (b) show results  for various $\lambda$ for dispersionless phonons $\delta=0$.  Panels (c) and (d) show results for various $\delta$ at strong coupling $\lambda=1$ corresponding to the unphysical regime. Inset of (d) shows the staggered displacement amplitude (Eq.~\eqref{eq:StagDisp}) as a function of $\delta$; the horizontal red dashed line at $1/2$ sets the  limit of physicality for the phonon displacements.}
 \label{fig:Rspace2}
 \end{figure}
 
In the linear approximation, $t(x_i - x_{j}) \approx  t - g\sqrt{2\Omega}(x_i-x_j)$, where $x_{i}\equiv\sqrt{\frac{1}{2\Omega}}(b_i^\dagger+b_i)$. This approximation is only valid for small distortions $\langle x_i-x_j\rangle \ll t/g\sqrt{2\Omega}$. As we show next, we always find evidence for lattice dimerization at the onset of  phase separation: $\langle x_i \rangle \sim (-1)^i \langle x\rangle$. Thus, here the model remains physical iff $|\langle b_i^\dagger+b_i\rangle|\ll \frac{t}{2g}=\frac{1}{2}\sqrt{\frac{2t}{\Omega\lambda}}$. We define the \emph{staggered} dimerization amplitude
\begin{align}
	\label{eq:StagDisp}
	{\cal D}&={1\over S}\sum\limits_{j=1}^{S}|\langle b_j^\dagger+b_j\rangle|,
\end{align}
computed over the \emph{typical} size $S\equiv S(\rho,\Omega,\lambda)$
of the dimerized region characterized by finite displacements  (width at half-maximum of the displacement profile) as a proxy for the viability of the linear model: results are physical only when ${\cal D} \sqrt{\Omega \lambda\over 2 t }< {1\over 2}$. The inset of Fig.~\ref{fig:PhaseDiagram} shows that this quantity is very small at small $\lambda$, but increases very sharply at larger $\lambda$. ${\cal D} \sqrt{\Omega \lambda\over 2 t}$ crosses the physicality limit of $1/2$ (gray dots in inset) marking the limits in the phase diagram (gray dots in main figure) beyond which the model become definitively unphysical due to the breakdown of the linear approximation.
This means phase separation exists only within a narrow sliver of the phase diagram whose width would be further reduced if a more stringent criterion ${\cal D} \sqrt{\Omega \lambda\over 2 t }\ll {1\over 2}$ is used. Unraveling the physics at very large electron-phonon couplings requires inclusion of nonlinear corrections to the coupling in the model, which is beyond the scope of this work.

Figure~\ref{fig:Rspace2} presents evidence that coalescing of bipolarons at $\lambda > \lambda_c$  represents a signature of true  phase separation. Figure~\ref{fig:Rspace2}(a)-(c) shows the average electron density $\langle \hat{n}_i\rangle$ as a function of position $i$ for a chain with OBC and $L=48$ sites and $N=6$ electrons (nominal density $\rho=0.125$) and $\Omega=2t$.  In the $\delta=0$ limit, depicted in  Fig.~\ref{fig:Rspace2}(a), we observe for $\lambda<\lambda_c$ a density profile with   $N/2$ peaks characteristic of a liquid of bipolarons; as $\lambda$ grows and crosses $\lambda_c$, the electrons clump into 
an electron-rich region forming a Gaussian-like droplet with a maximum density $n=0.5$, surrounded by electron-free regions, a behavior characteristic of phase separation~\cite{Moreno2011}.  We emphasize that due to the use of OBC, the electron-rich, 
phase-separated region is centered at $L/2$.  Simulations with PBC (see 
Supplemental Materials~\cite{SupplementaryMaterials}) give identical results at 
strong coupling, but with the center of the electron-rich region not pinned to a specific site. 
Figure~\ref{fig:Rspace2}(c) establishes that even at strong coupling,
{\it e.g.} $\lambda=1$, a sufficiently large $\delta$ melts the electron-rich region 
re-instating the bipolaronic liquid phase. Figure~\ref{fig:Rspace2}(b)-(d) 
demonstrates that the lattice indeed dimerizes: 
$(-1)^i\langle b_i+b_i^\dagger\rangle>0$ in the core of the electron-rich region; 
for $\lambda>\lambda_c$. In particular, Fig.~\ref{fig:Rspace2}(b) shows that 
for $\lambda=1>\lambda_{c}$ the phonon displacement amplitude within 
the electron-rich region far exceeds the physically allowed limit 
${\cal D}\sqrt{\Omega\lambda/2t}=1/2$. Fig.~\ref{fig:Rspace2}(d) shows 
that $\delta$ can be \emph{fine-tuned} so that an electron-rich region 
is accompanied by lattice dimerization within the stability limits of the model. 
We have also verified (not shown) that double occupancy in the core of 
the electron-rich region is very small 
$\langle \hat{n}_{i\uparrow} \hat{n}_{i\downarrow} \rangle\simeq 0.125$, 
which implies that a moderate repulsive Hubbard $U$ 
will not affect these results significantly, see, {\it e.g.}, \cite{CoulombPeierls}.

{\em Small but finite electron concentrations.---} Having established the stability of bipolarons in the extremely dilute limit, we also confirm the stability of the bipolaron liquid for finite densities $\rho=0.125-0.5$. We construct the phase diagram as a function of $\rho$ in 
Fig.~\ref{fig:PhaseDiagram2} for $\delta=0.04$.  We find, as before, a robust bipolaronic liquid that is stable up to very large values of $\lambda$, almost everywhere the linear approximation is valid.   The critical value $\lambda_c$ above which phase separation is energetically favorable decreases with electronic filling. In the Supplemental Materials~\cite{SupplementaryMaterials}, we determine the critical $\delta$ at strong coupling $\lambda=1$ such that 
the dimerized phonon displacement amplitude falls within the physical limit, finding that  $\delta_c$ increases nearly linearly from the dilute density limit up to quarter filling.  These results conclusively demonstrate the robustness of the bipolaron liquid phase in the Peierls model \emph{beyond} the extremely dilute density limit. This is particularly relevant for commensurate densities such as quarter filling $\rho=0.5$, where  instabilities to orders other than phase separation might have been anticipated.

{\em Conclusions.---} We have numerically studied a model with linear Peierls electron-phonon coupling on a 1D chain, and proved that a dilute bipolaron liquid is stable against phase separation up to large couplings $\lambda < \lambda_c$. For $\lambda > \lambda_c$ we present evidence for a new type of phase separation, with several interesting properties including lattice dimerization within the electron-rich region surrounded by electron poor, undimerized regions. The region of validity of the linear approximation terminates close to $\lambda_c$, so that whether phase separation occurs in the model upon inclusion of  higher-order, non-linear couplings, remains an open question.  Furthermore, we have confirmed the stability of bipolaron liquids in most of the physical parameter space for small carrier densities extending up to quarter filling.  In contrast, in more studied models of electron-phonon coupling like in the extended Holstein model,  phase separation sets in at a much smaller $\lambda_c$.
These results reinforce the possibility of bipolaronic high-temperature superconductivity in  models with Peierls-like coupling  
at low carrier densities, a  scenario  believed to be impossible for other forms of electron-phonon coupling~\cite{Ranninger1998}.

{\em Acknowledgements.}---  We acknowledge useful discussions with V.~Cataudella, G.~De Filippis, S.~Johnston, S.~Li,  A.~J. Millis, and C.~A. Perroni. A.~N. acknowledges computational resources and services provided by Compute Canada and Advanced Research Computing at the University of British Columbia.   A.~N. and M.~B. acknowledge support from the Max Planck-UBC-UTokyo Center for Quantum Materials, the Canada First Research Excellence Fund (CFREF) Quantum Materials and Future Technologies Program of the Stewart Blusson Quantum Matter Institute (SBQMI), and the Natural Sciences and Engineering Research Council  of Canada (NSERC). J.~S. acknowledges support from the National Science Foundation (NSF) Materials Research Science and Engineering Centers (MRSEC) program through Columbia University in the Center for Precision Assembly of Superstratic and Superatomic Solids under Grant No.~DMR-1420634 and the hospitality of the Center for Computational Quantum Physics (CCQ) at the Flatiron Institute. The Flatiron Institute is a division of the Simons Foundation. A.~E.~F. acknowledges support from NSF grant No.~DMR-1807814.

%

\clearpage
\end{document}


\clearpage
\renewcommand\thefigure{\thesection{S}\arabic{figure}}    
\setcounter{figure}{0} 

\onecolumngrid

\section{\Large S\lowercase{upplementary} M\lowercase{aterials for:}\\ B\lowercase{ipolaron liquids at strong Peierls electron-phonon couplings}
}

\author{Alberto Nocera}
\affiliation{Department of Physics and Astronomy, University of British Columbia, Vancouver, British Columbia, V6T 1Z1 Canada}
\affiliation{Stewart Blusson Quantum Matter Institute, University of British Columbia, Vancouver, British Columbia, V6T 1Z4 Canada}

\author{John Sous}
\affiliation{Department of Physics, Columbia University, New York, New York 10027, USA}

\author{Adrian E. Feiguin}
\affiliation{Department of Physics, Northeastern University, Boston, Massachusetts 02115, USA}

\author{Mona Berciu}
\affiliation{Department of Physics and Astronomy, University of British Columbia, Vancouver, British Columbia, V6T 1Z1 Canada}
\affiliation{Stewart Blusson Quantum Matter Institute, University of British Columbia, Vancouver, British Columbia, V6T 1Z4 Canada}

\date{\today}
\maketitle
\normalsize

\renewcommand\thefigure{\thesection{S}\arabic{figure}}    
\setcounter{figure}{0} 

\onecolumngrid
\section{Details of density matrix renormalization group (DMRG) simulations:\\
OBC versus PBC results}
Here we discuss additional computational details about the DMRG simulations performed in this work. 

\begin{figure}[!h]
\includegraphics[width=0.8\columnwidth]{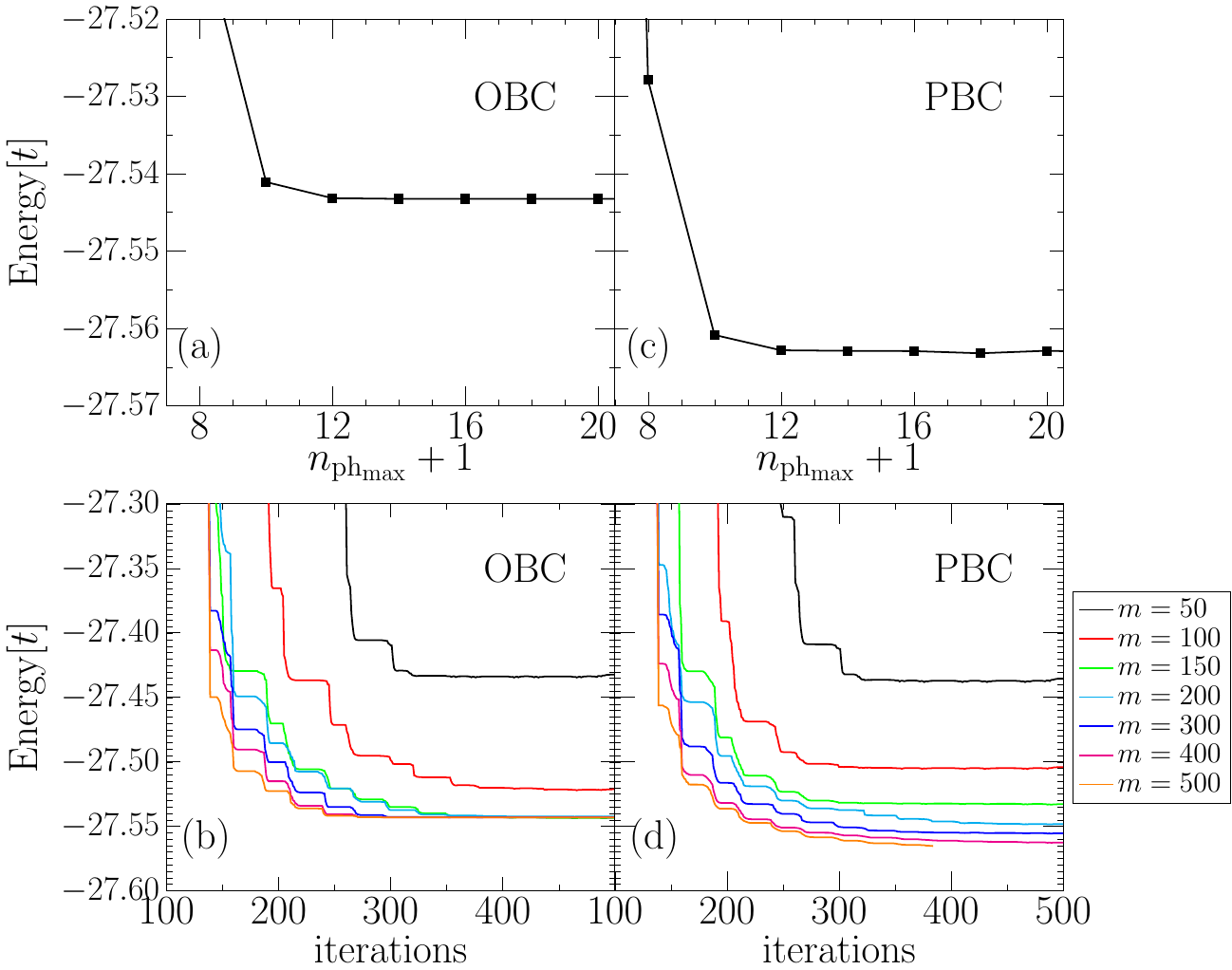}
\caption{(color online) {Convergence of the ground-state energy 
 obtained using DMRG with respect to the local Hilbert space dimension 
$n_{\rm ph_{\rm max}} + 1$ (top) and number of iterations (bottom) for the 1D Peierls 
model with open (OBC) (left) and periodic (PBC) (right) boundary conditions.} 
In this figure, we study an $L=16$ chain for $\lambda=1$ and $\Omega=2t$.
}\label{fig:S1}
\end{figure}

Figure~\ref{fig:S1}(a)-(c) shows the ground-state energy as a function of the number of states kept in the local phonon Hilbert space for a $L=16$ chain, with $\lambda=1$ and $\Omega=2t$. The data shows that in the strong coupling regime the ground-state energy converges for $n_{{\rm ph}_{\rm max}}+1\simeq 12$ with a precision of $10^{-4}$ for OBC and $10^{-3}$ for PBC. As already mentioned in the main text, a bond dimension $m=600$ suffices to obtain convergence with a truncation error smaller than $5\times 10^{-7}$ for OBC and $5\times 10^{-6}$ for PBC for chains with length $L\leq48$ and with $N\leq16$ electrons. 

Figure~\ref{fig:S1}(b)-(d) confirms a known result to DMRG practitioners: numerical convergence proves to be much harder
for PBC than for OBC. The data shows that even for a small system ($L=16$ sites) up to 40 DMRG sweeps (with $L-1$ iterations per sweep) are needed to numerically converge to the precision set above. For the $L=48$ data shown in the main text, we performed up to 100 DMRG sweeps.

\begin{figure}[!h]
\includegraphics[width=0.8\columnwidth]{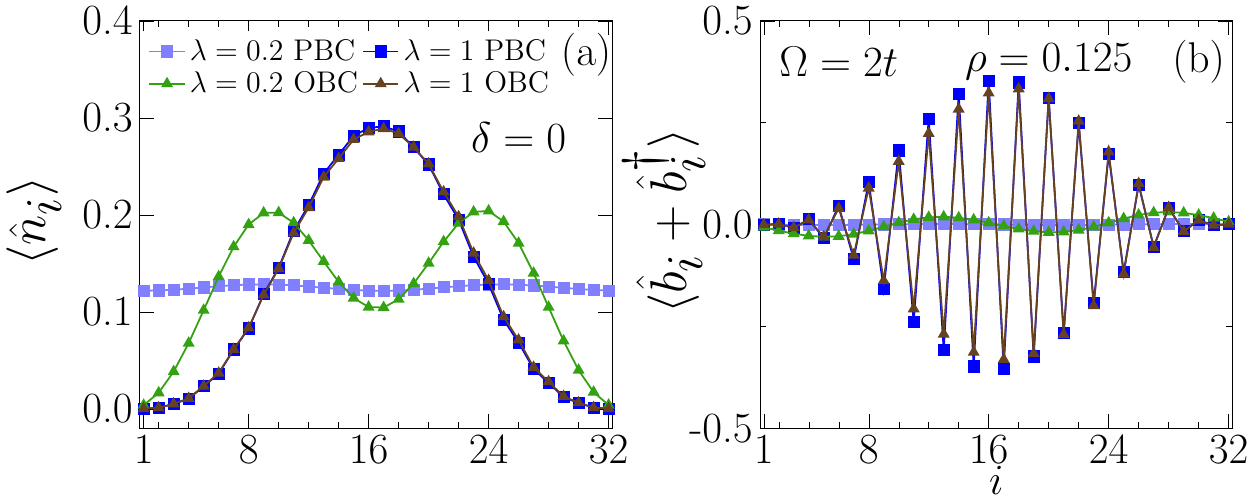}
\caption{(color online) GS expectation values of the  average occupation number 
	  $\langle \hat{n}_i\rangle $ (a) and lattice distortion $\langle b_i+b_i^\dagger\rangle$ (b) for the 1D Peierls 
model with periodic (blue square symbols) and open (green triangle symbols) boundary conditions.
In this figure, we study an $L=32$ chain with $\Omega=2t$ at both weak coupling $\lambda=0.2$ and strong coupling $\lambda=1$.}\label{fig:S1A}
\end{figure}

We note that a precise determination of the spatial electronic density profile in the strong coupling
regime would be generally numerically challenging for the following reasons. Many almost-degenerate states very 
close to the ground state connected by translation of the electron-rich region with respect to the electron-poor 
region exist. This is compounded by the difficulty associated with the study of very large systems 
with sufficiently large number of electrons (whilst maintaining the density $n$ extremely small but finite) 
needed to accurately represent a \emph{phase} in the thermodynamical limit \cite{Moreno2011}.
To further clarify this point, Fig.~\ref{fig:S1A} shows very good convergence of the density and 
lattice distortion profiles when comparing OBC against PBC. 
Indeed, at weak coupling $\lambda=0.2$, 
the charge density profile is uniform for PBC, while
it shows typical Friedel-like density oscillations for OBC with the number of peaks being proportional to the number of electron pairs, or \emph{bipolarons}. In this regime, the lattice distortion profile is almost zero and it is negligible.
At strong coupling $\lambda=1$, however, we observe an almost perfect agreement 
between OBC and PBC for both the density and lattice dimerization profiles, 
convincingly showing the validity of our results in the thermodynamic limit. On the 
other end, since in the presence of PBC the Hamiltonian is translationally 
invariant, the agreement between OBC and PBC can be viewed as 
an artifact of DMRG which is unable to fully restore 
the translational invariance of the model due to the tendency towards phase 
separation.

\section{Details of the phase diagram of the Peierls model}
\subsection{Dilute electron density}
\begin{figure}[!h] 
\includegraphics[width=0.9\columnwidth]{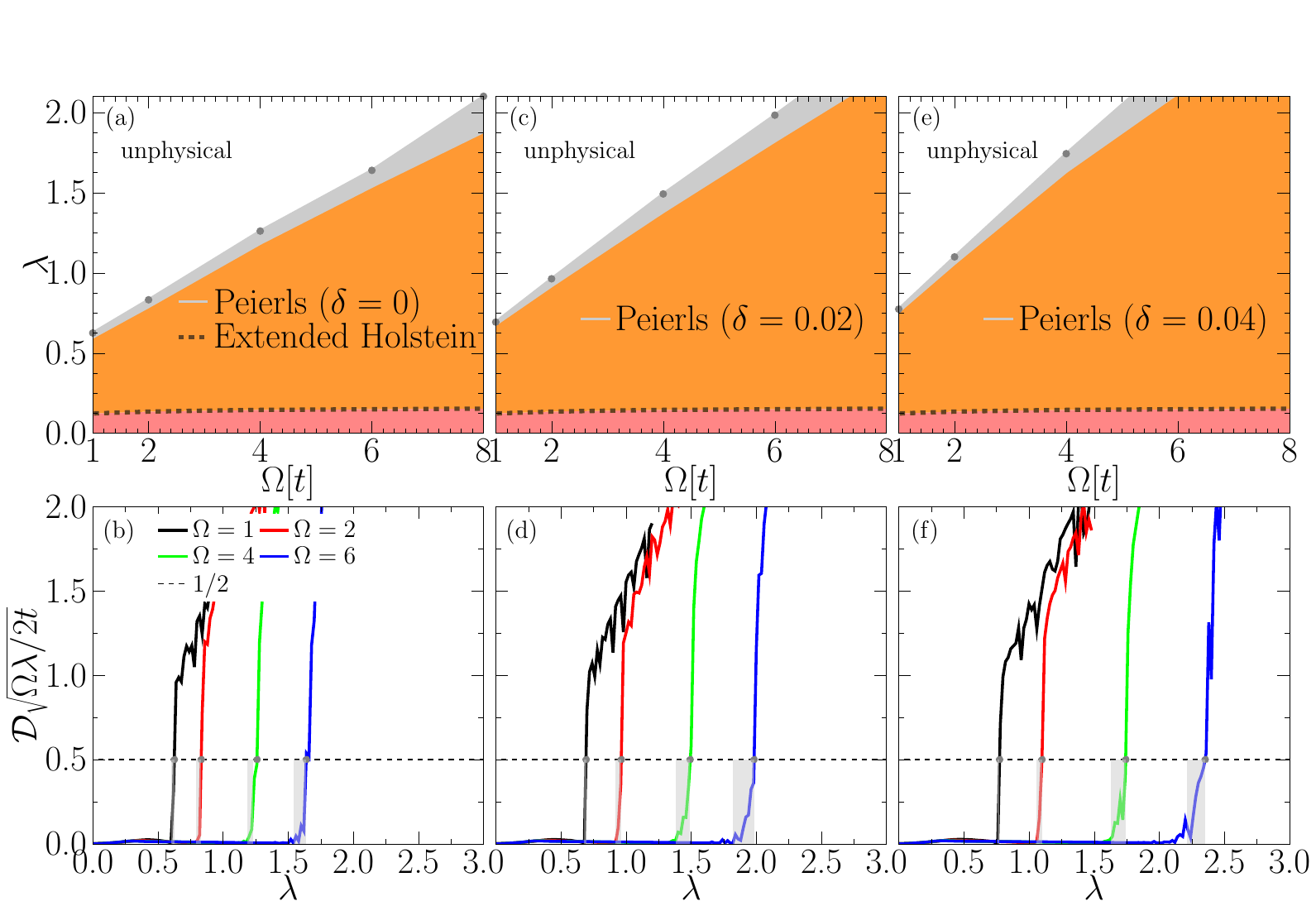}
\caption{(color online) Panels (a), (c) and (e): Phase diagrams of the 1D Peierls model
	  contrasted against that of  the extended Holstein (EH) model in the dilute electron density limit for various  $\delta$. Panels (b), (d) and (f): Staggered displacement amplitude ${\cal D}$, 
	  defined in Eq.~(\ref{eq:StagDisp}), as a function of $\lambda$ for different values 
	  of $\Omega$. 
	  The horizontal dashed line at 1/2 defines the limit beyond which phonon displacements become unphysical. 
	  As in Fig.~\ref{fig:PhaseDiagram} in the main text, 
	  these results were obtained using DMRG for a system with $L=32$ sites for $N=6$ electrons.
	  }\label{fig:S2}
\end{figure}

In this subsection, we study the phase diagram of the model in the dilute carrier density limit as a function of
the parameter $\delta$, which gives 
a dispersion for the optical phonons. The interaction term in the Hamiltonian of the Peierls model is by definition 
linear in the relative displacement of neighboring sites across a bond.  Large breathing distortions in this model may result in situations in which the bond length becomes negative with a consequent change of sign of the hopping, an indication of the breakdown of the linear approximation. The $\delta$ term in the phonon dispersion disfavors this unphysical behavior. Indeed, Fig.~\ref{fig:S2} shows that it energetically 
penalizes large staggered displacement amplitudes, and importantly renders the bipolaronic liquid phase region of the phase diagram larger.

\subsection{Finite electronic density}
\begin{figure}[!h]
\hspace{-0.35cm}\includegraphics[width=0.505\columnwidth]{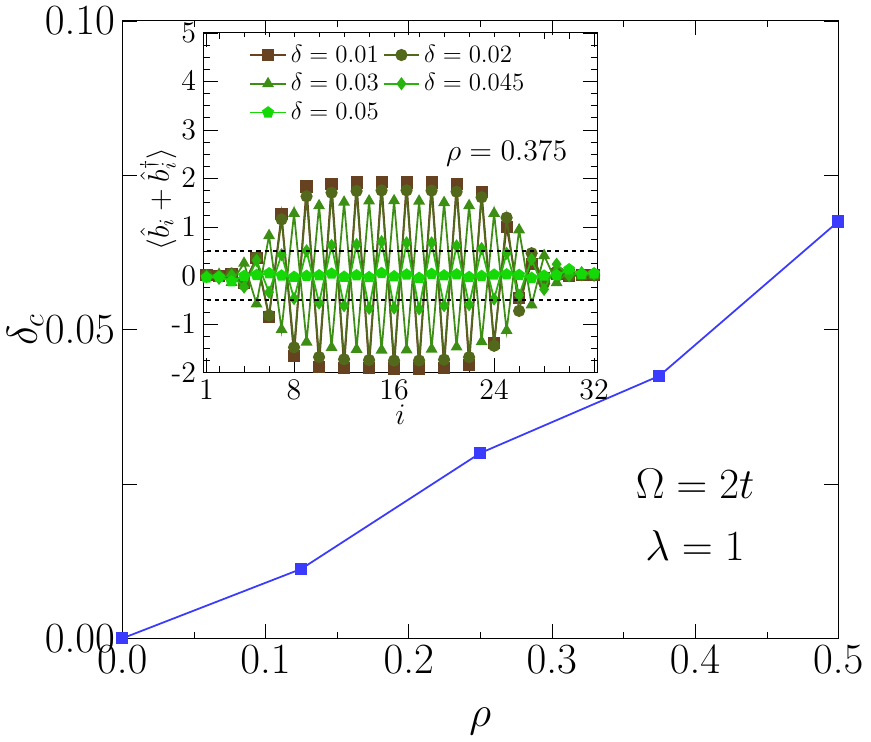}
	  \caption{(color online) Main: Minimum value of $\delta$, denoted as $\delta_c$, required such that the displacements $\langle x_i \rangle$ fall within the physical 
	  regime (${\cal D}\sqrt{\Omega\lambda/2t}<1/2$),  as a function 
      of the density of electrons $\rho$.
	  Inset: GS expectation values of the  lattice distortion $\langle b_i+b_i^\dagger\rangle$ as a function of the lattice site $i$.
          This result was obtained for a chain with $L=32$ and OBC for $N=12$ electrons ($\rho=0.375$) 
          at phonon frequency $\Omega=2t$ and $\lambda=1$. }
 \label{fig:Sdeltac}
 \end{figure}

Here, we supplement the phase diagram of the model constructed in $\lambda$-$\rho$ space (Fig.~\ref{fig:PhaseDiagram2} of the main text), and determine in Fig.~\ref{fig:Sdeltac} the minimal value of the phonon dispersion amplitude $\delta_c$  
such that the dimerized phonon displacement amplitude falls within the physical limit ${\cal D}\sqrt{\Omega\lambda/2t} <1/2$ (see main text and Eq.~\eqref{eq:StagDisp} for the definition of ${\cal D}$). We find that $\delta_c$ increases nearly linearly from the dilute density limit up to quarter filling (main panel of Fig.~\ref{fig:Sdeltac}). The inset of Fig.~\ref{fig:Sdeltac} explains the procedure we use to numerically extract the value of $\delta_c$ for a given $\rho$.  We finally note that we have verified that the phonon dispersion
parameter $\delta$ has no effects on computed observables for $\delta\leq 0.05$.  These results establish the robustness of the bipolaron liquid phase 
in the Peierls model in the small -- but finite -- electron density limit.

\section{Stability of bipolaronic phase:\\
Maxwell construction}
In this section we use the Maxwell construction to assess the stability of the bipolaronic phase and compare the Peierls, Fr\"ohlich, and the Hubbard model with no electron-phonon coupling. The procedure of the Maxwell construction
consists in finding the value of chemical potential and electronic concentration $(\mu=\mu^*,n=N_{\rm tot})$
that minimizes the grand potential function  
\begin{equation}\label{eq:Grand}
G(\lambda,\mu,n)=E_n-\mu n.    
\end{equation}
Here, $E_n \equiv E(n=N_{\rm tot},s=S_z)$ is
the ground state energy of the model 
on a finite chain with $n=N_{\rm tot}$ electrons and spin projection $s=S_z$. 
For even $N_{\rm tot}$, the ground state lies in the 
 $s=S_z=0$ sector, while for odd $N_{\rm tot}$, it lies in the  $s=S_z=1/2$ sector. We have also verified that when $N_{\rm tot}=n$ is odd, $E(n,s)=E(n,-s)$ as a result of time reversal symmetry.  To lighten the notation we will omit the spin quantum number of the ground state in what follows.

\begin{figure}[!h]
\includegraphics[width=0.9\columnwidth]{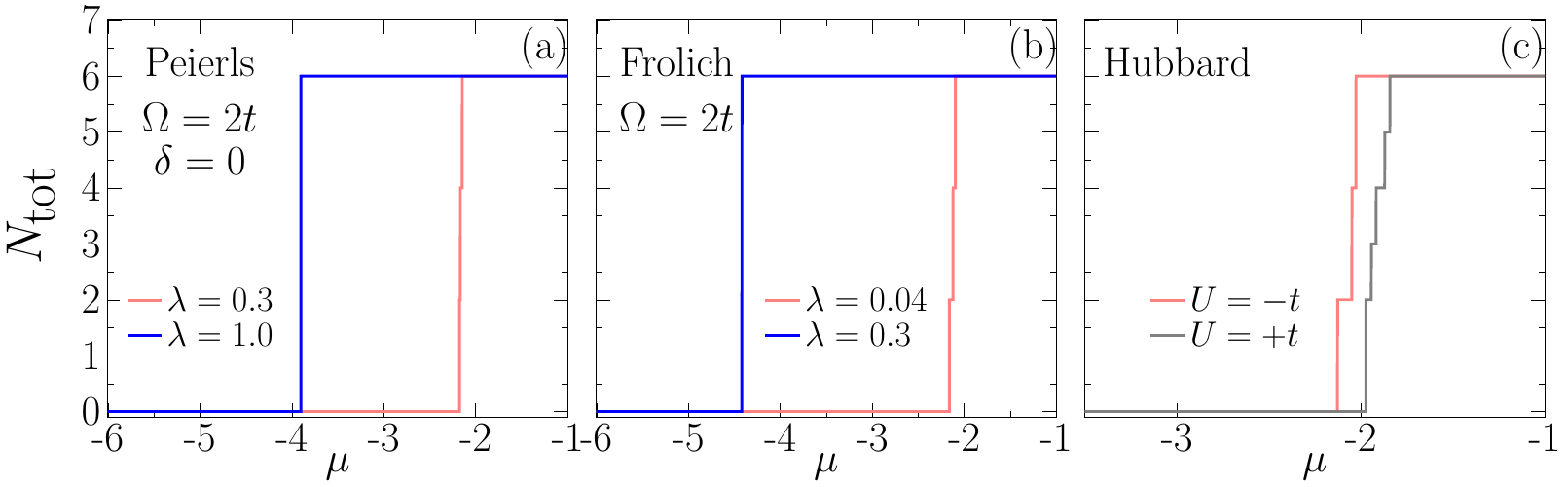}
\caption{(color online) Maxwell construction: Total number of electrons $N_{\rm tot}$ that minimizes the grand potential function in Eq.~\eqref{eq:Grand} as a function of the chemical potential $\mu$ (see text for further details) for the Peierls (a), Fr\"ohlich (b) and Hubbard (with no electron-phonon coupling) (c) models. The results were obtained for a chain with $L=32$ for up to 6 electrons, at   $\Omega=2t$ and $\delta=0$.}\label{fig:S4}
\end{figure}

Figure~\ref{fig:S4} shows the value of  $N_{\rm tot}$ that minmizes the grand potential $G$ 
as a function of the chemical potential $\mu$ and electron-phonon 
coupling strength $\lambda$ for the Peierls (a), Fr\"ohlich (b) and Hubbard  (c) models.
We show representative points in the weak- and strong- coupling regime of the Peierls and  Fr\"ohlich models,
while comparing the cases of repulsive and attractive interaction strengths in the Hubbard model. The first interesting feature revealed in this procedure is that electronic 
concentrations associated with \emph{odd} $N_{\rm tot}$ (greater than one)
never minimize $G$ for any value of $\mu$ and $\lambda$ 
(or electron-electron interaction $U$ in the case of the Hubbard model). 
In fact, for sufficiently small $\lambda$, as the chemical potential is increased, $N_{\rm tot}$ that minimize $G$ \emph{jumps} first 
from a value 0 to 2, then from 2 to 4, 
and finally from 4 to 6. We remark that finite jumps are 
observed because our numerical simulations are carried out in a finite system and thus correspond to a  fixed value of electronic density. This implies that bound states
made of one, three and five electrons are not \emph{stable} 
solutions of the Peierls and Fr\"ohlich models at low electronic 
concentrations (Fig.~\ref{fig:S4}(a)-(b)).
The presence of jumps of two units (from 0 to 2,  2 to 4, and from 4 to 6)
indicate that, at least for $\lambda$ smaller than a certain threshold,
the ground state is made of two-electron bound states, 
that is \emph{bipolaron}  states.
For sufficiently large 
$\lambda$ in both Peierls and Fr\"ohlich models we observe a 
finite jump of $N_{\rm tot}$ from 0 
to 6. This indicates that bound states with $0<N\leq 6$ or larger  
can coexist in the system, indicating phase 
separation. For a  comparison, Fig.~\ref{fig:S4}(c) shows how the Maxwell construction 
works in a Hubbard chain with attractive as well as repulsive onsite 
Hubbard $U$ interaction. For attractive $U$, we observe the presence of jumps in units of two 
 (from 0 to 2,  2 to 4, and from 4 to 6) in the total density, highlighting 
clustering of electrons in pairs. For weakly repulsive 
$U$, electrons repel each other,
pairs are not formed, and jumps in units of one in the total density are found instead.

\begin{figure}[!h]
\includegraphics[width=0.5\columnwidth]{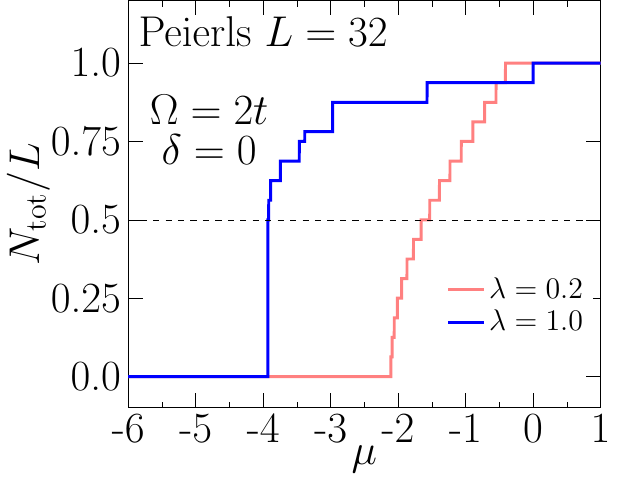}
\caption{(color online) Particle density (total number of electrons $N_{\rm tot}$ per site) that minimizes the grand potential function in Eq.~\eqref{eq:Grand} as a function of the chemical potential $\mu$ (see text for further details) for the Peierls model at finite electron densities up to half filling $\rho=1$. These results were obtained for a chain with $L=32$, at $\Omega=2t$ and $\delta = 0$. The horizontal dashed line at $N_{\rm tot}/L=0.5$ serves as a guide to the eye.}\label{fig:S6}
\end{figure}

Finally, Fig.~\ref{fig:S6} shows Maxwell construction results for the Peierls model
at finite electron densities up to half filling. 
It is clear that at weak coupling $N_{\rm tot}/L$ increases in step of two electrons from $N_{\rm tot}/L=0$ to $N_{\rm tot}/L=1$, indicating a bipolaronic liquid phase.
At strong coupling (where the phonon distortions are very large and unphysical, see main text) 
the Maxwell construction unveils the unphysical phase-separated phase between 
$N_{\rm tot}/L=0$ and 
quarter-filled states $N_{\rm tot}/L=0.5$. Based on this, we argue that in the thermodynamic limit, the state $N_{\rm tot}/L=0.5$ will be stable only for $\mu=\mu^{*}\simeq-4$.
Indeed, at the jump $\mu=\mu^{*}\simeq-4$, the derivative 
$\partial N_{\rm tot}/ \partial\mu|_{\mu=(\mu^{*})^{+}}\rightarrow\infty$, while  for $\mu>\mu^{*}$ the density increases in steps of two electrons as a function of $\mu$ up to $N_{\rm tot}/L=0.75$.  

%

\clearpage